\documentstyle[aps,prl,multicol,epsf]{revtex}
\title{\bf Aging Exponents in Self-Organized Criticality}

\author{Stefan Boettcher$^{1,2}$}
\address{
$^1$Center for Theoretical Studies of Physical Systems, Clark Atlanta
University, Atlanta, GA 30314\\
$^2$Center for Nonlinear Studies,
MS-B258, Los Alamos National Laboratory, Los Alamos, NM 87545}
\date{\today}
\begin{document}
\maketitle\parskip 2ex

\begin{abstract}
In a recent Letter [Phys. Rev. Lett. {\bf 79}, 889 (1997)]
we have demonstrated that the avalanches in the Bak-Sneppen
model display aging behavior similar to glassy systems.  Numerical
results for temporal correlations show a broad distribution with two distinct
regimes separated by a time scale which is related to the age
of the avalanche.  This dynamical breaking of time-translational invariance
results in a new critical exponent, $r$. Here we present results for
$r$ from extensive numerical simulations of self-organized critical
models in $d=1$ and $2$. We find $r_{d=1}=0.45\pm 0.05$ and
$r_{d=2}=0.23\pm 0.05$ for the Bak-Sneppen model, and our results suggest 
$r=1/4$ for the analytically tractable multi-trade model in
both dimensions. 

\end{abstract}

\begin{multicols}{2}

\section{Introduction}

Self-organized criticality (SOC) \cite{BTW} describes a general
property of slowly-driven dissipative systems with many degrees of
freedom to evolve intermittently in terms of bursts spanning all
scales up to the system size.  Many natural avalanche-like phenomena
have been represented using this concept, including earthquakes
\cite{G+R,earthquake,Ito}, extinction events in biological evolution
\cite{Raup,G+E,B+S}, and landscape formation \cite{land,per}.  Recently,
SOC has been observed in controlled laboratory experiments on rice
piles \cite{Frette}.  Theoretical models of rice piles \cite{Oslo} are
related to a variety of different physical systems by universality
\cite{PaBo}.  One crucial ingredient for a system to exhibit SOC is
the existence of thresholds that allow it to record the stress exerted
by the driving force over long periods of time.  The emergence of
long-term memory has been demonstrated analytically
\cite{BoPa2} for a multi-trait evolution model
\cite{BoPa1}, a variant of the Bak-Sneppen model \cite{B+S}.

In a recent Letter \cite{BoPa3} we have shown that the self-organized
critical state in the Bak-Sneppen model exhibits aging behavior that
is reminiscent of glassy systems \cite{slowdyn}.  Our results indicate
that intrinsic two-time autocorrelation functions $P(t_w;t)$, describing 
the return of activity to a site at time $t_a=t+t_w$ which was active most
recently at time $t_a=t_w$ for an avalanche that started at $t_a=0$,
decay as power laws with two distinct regimes according to
\begin{eqnarray}
P(t_w; t)\sim  t^{-\tau_{\rm first}} &
 f\left({ t \over t_w}\right)&\cr
\noalign{\medskip}
&f(x)&\sim\cases{ {\rm constant} &$(x\ll 1)$,\cr
\noalign{\medskip}
x^{-r} &$(x\gg 1)$.}
\label{pfirst}
\end{eqnarray}
The early time regime is that of the
familiar stationary dynamics.  The late time regime has a new
 critical coefficient $r$ characterizing the nonstationary relaxation
behavior of the SOC systems.  The ``waiting time'' $t_w$ separating the early
and late time regimes is a measure of the age of the avalanche.  We have argued
that this aging behavior arises from the hierarchical structure of the
avalanches.

Generally, the origin of aging behavior does not have to be
profound. For instance, it arises in a simple random walk model near
a wall, where a symmetry (translational invariance) is explicitly
broken.  But while the
random walk near a wall and the Bak-Sneppen model (especially its
random neighbor variant \cite{DeB}) are similar in many ways, it
appears that those similarities do not explain the observed aging behavior.

In Sec. II, we use the random walk near a wall to illustrate the
quantities that will be measured for the SOC models. In Sec. III, we
consider the important technical issue of extracting the intrinsic
aging behavior for a process that does not conserve its norm (such as a
random walk near an absorbing wall or a SOC avalanche, having a finite 
stopping probability).
In Sec. IV, we present detailed numerical results for the
aging behavior of SOC models in one and two dimensions. In particular,
we have simulated the Bak-Sneppen model and the multi-trade model. 
We have also simulated sandpile models exhibiting SOC which show a
quite different behavior and will be discussed elsewhere \cite{BoPa4}. 
In Section V we discuss our results and show that a simple random walk 
description is not sufficient to explain the found aging behavior.

\section{Aging Random Walks}

In this section we will discuss a random walk model to provide
a simple intuitive picture of aging. It shows how the
system memorizes 
the (here, explicit) breaking of a symmetry. Furthermore, the random
walk illustrates the meaning of the correlation
functions used to describe the aging behavior and serves to discuss
some of the technical issues in measuring those functions.

We want to consider a random walker on a $d=1$ lattice who can jump at
most one step on each update either on the infinite lattice or a
semi-infinite lattice with an absorbing wall at the origin. A random
walk is completely 
described  by its propagator, the conditional probability
$G(n,t|n_0,t_0)$ for a walker to reach a site $n$ at time $t$, given
that it was at site $n_0$ at some previous time $t_0<t$. To determine
its aging behavior we want to compute a simple two-time correlation 
function (see e. g. Ref.~\cite{S+H} for a similar definition)
\begin{eqnarray}
P(t_w;t)=\sum_n G(n,t+t_w|n,t_w)~ G(n,t_w|n_0,0)
\label{twotime}
\end{eqnarray}
for a walker to return to a site at time $t_a=t+t_w$, given that it was
at the same site at the ``waiting time'' $t_w$ after the start of the
walk. Thus, to determine the return
probability to a site, we take the time $t_w$ from the start of the
walk to a previous passage at that site into account. If the two-time
correlation function $P$ explicitly depends on $t_w$, the system is
said to ``age'' because the walk would retain a memory of the time
since its inception in form of a return probability that evolves over time. 

\subsection{Unconstrained Random Walks}
\label{uwalk}

Clearly, for an unconstraint walk, $G$ is invariant in space and time,
and it is $G(n,t+t_w|n,t_w)=G(0,t|0,0)$. Since the norm of the walk is
preserved at all times, $\sum_n G(n,t_w|n_0,0)\equiv 1$,
Eq.~(\ref{twotime}) gives $P(t_w;t)\equiv P(t)\equiv G(0,t|0,0)$,
independent of $t_w$. Thus the unconstrained walk has no memory of its
past and does not age.

\subsection{Random Walks near an Absorbing Wall}

In the presense of an absorbing wall at the origin, spatial invariance
is explicitly broken while time invariance for $G$ still holds. 
Eq.~(\ref{twotime}) merely simplifies to
\begin{eqnarray}
P(t_w;t)=\sum_{n>0} G(n,t|n,0)~ G(n,t_w|n_0,0)
\label{newc}
\end{eqnarray}
which remains dependent $t_w$. Hence the breaking of spatial
invariance in $G$ leads to a breaking of time invariance in $P$. Such
a memory effect arises in the following manner: While
in the unconstrained case the mean distance $<n>$
of the walker from its origin vanishes, the walker starting near a
wall departs from it such that $<n>\sim
t^{1/2}$, which follows from the propagator given by
\begin{eqnarray} 
G(n,t|n_0,0)\approx {1\over \sqrt{\pi
t}}\left[e^{-{(n-n_0)^2\over 4t}}- e^{-{(n+n_0)^2\over 4t}}\right]
\label{result}
\end{eqnarray}
for sufficiently large $t$ and $n$. Since the distribution for all
walks is sharply peaked near its mean, most walkers occupy sites
$n\sim t_w^{1/2}$ away from the wall after the waiting time $t_w$. {\it
Given} that most walkers occupy such sites $n$, returns to that site
during subsequent times $t\ll t_w$ follow the statistics of an
unconstrained walk: $t$ is not yet large enough to return to the wall. 
Only after times $t\sim t_w$ do a sizable portion of the walkers
experience the effect of the distant wall again, which leads to a
change in their return statistics for all times $t\gg t_w$. The
change-over in its return behavior at later times $t+t_w$ thus
provides the walker with a 
memory of the earlier period from the start up to time $t_w$ when the
walk was drifting away from the wall. (Considering the walker as the
center of a growing domain and its distance to the wall as a measure
of the linear size of that domain relates this random walk model
nicely to the domain-growth picture of aging in glasses proposed in
Ref.~\cite{K+H}.)  

{}For this random walk model the crossover in its return behavior can
be easily derived explicitly: Inserting the appropriate forms of $G$ in
Eq.~(\ref{result}) into Eq.~(\ref{newc}) and choosing $n_0$ arbitrarily
close to the wall, we obtain asymptotically 
\begin{eqnarray}
P(t_w;t)&\approx&{n_0\over\pi\sqrt{t\,t_w^3}}\int_0^{\infty}dn
\left[1-e^{-{n^2\over t}}\right] n e^{-{n^2\over 4t_w}}\cr
\noalign{\medskip}
&=&{n_0\over \sqrt{t_w}}{2\over \pi} \,t^{-{1\over 2}} \,
f\left({t\over t_w}\right),\quad f(x)={1\over 1+{x\over 4 }}.
\label{cabso}
\end{eqnarray}
Thus, time-translational invariance is broken  and the walk appears to
age because $P$ becomes a function of the scaling variable $x=t/t_w$ 
signaling the predicted crossover in the return behavior that
scales linearly with the waiting time $t_w$.
Here the scaling function $f(t/t_w)$ behaves  asymptotically as
$f(x\ll 1)\sim 1$ and $f(x\gg 1)\sim 4/x$, but to obtain the
``intrinsic'' aging behavior of this process, we have to consider the
effect that the norm is not preserved because at each time step walkers
may disappear at the absorbing wall. 

\section{Intrinsic vs. Measured Aging Behavior}

Consider the unconstrained random walk in Sec. \ref{uwalk}, but with a
finite probability $z$ to disappear at each time step. Then, the
propagator $G_z$ of this walk is given by $G_z(n,t|n_0,t_0)=(1-z)^{t-t_0}
G(n,t|n_0,t_0)$, where $G$ is the propagator of the norm-preserving walk
in Sec. \ref{uwalk}. Thus, with $\sum_n G(n,t|n_0,t_0)=1$, we get $\sum_n
G_z(n,t|n_0,t_0)=(1-z)^{t-t_0}$, i. e. the norm of this process is not
preserved. According to our definition of the two-time auto-correlation
function $P$ in Eq.~(\ref{twotime})
\begin{eqnarray}
P^{\rm meas}(t_w;t)&=&\sum_n  G_z(n,t+t_w|n,t_w)~ G_z(n,t_w|n_0,0)\cr
&=& G(0,t|0,0) (1-z)^{t+t_w},
\end{eqnarray}
we would have to conclude that this process ages, since $P$ depends on $t_w$.
And, indeed, in a numerical simulation of the process we would measure
$P^{\rm meas}$, because we  would average over all processes up to a
temporal cut-off $t_{\rm co}$, including those that disappear at times
$t<t_{\rm co}$.

But, clearly, the ``aging'' in this simple process is an artifact due to
the diminished norm $(1-z)^{t+t_w}$ at time $t+t_w$. Thus, proper
normalization is required to extract the ``intrinsic'' aging behavior 
(due to the infinite walk) from the ``measured'' aging
behavior\cite{star}.

Other process, such as random walks near an absorbing wall or avalanches 
in the Bak-Sneppen mechanism below, also may disappear before reaching the 
cut-off, and we have to consider the effect on the statistics of the 
measured results. Here, too, one is
interested in the intrinsic properties of the surviving process,
i. e. those of the infinite random walk or avalanche, while in
simulations one usually averages over all runs of a process, whether
they survived or not. In many cases, the results for the asymptotic
scaling behavior of some intrinsic property are no different than the
measured ones because the contribution from dying runs remains
insignificant. But for the correlation functions considered in this
paper which depend on two independent time variables, we do need to
consider the effect of the probability $P_t(\theta)$ of a process to
disappear at time $\theta$ to relate the intrinsic and the measure results.

To obtain the {\it intrinsic} properties
of the infinite process, we have to properly normalize 
the correlation function. To that end, we consider the two-time correlation 
function $P(t_w;t|\theta)$ for a run that disappears exactly at time
$t_a=\theta$, and its generic relation to the 
intrinsic two-time correlation function $P^{\rm intr}(t_w;t)$:
\begin{eqnarray}
P(t_w;t|\theta)=\cases{0 &$(\theta<t_w+ t)$,\cr
\noalign{\medskip}
P^{\rm intr}(t_w;t)  &$(\theta\geq t_w+ t)$.}
\end{eqnarray}
These quantities are related to the measured two-time correlation function 
$P=P^{\rm meas}$ given in Eq.~(\ref{cabso}): Assuming a power-law
probability $P_t(\theta)\sim \theta^{-\tau}$, $\tau>1$, for the 
run to disappear at time $t_a=\theta$, we have 
\begin{eqnarray}
P^{\rm meas}(t_w;t)&=&\int_0^{t_{\rm co}} d\theta~P(t_w;t|\theta)~
P_t(\theta)\cr
\noalign{\medskip}
&\sim& P^{\rm intr}(t_w;t)\left[\left(t_w+ t\right)^{1-\tau}-t_{\rm co}^{1-\tau}\right].
\end{eqnarray}
Assuming that we only consider data sufficiently far from the cut-off,
i. e. $\left(t_w+ t\right)^{1-\tau}\ll t_{\rm co}^{1-\tau}$, we obtain
\begin{eqnarray}
P^{\rm intr}(t_w;t)&\sim& P^{\rm meas}(t_w;t)~\left(t_w+ t\right)^{\tau-1}.
\end{eqnarray}
{}For the particular form of the intrinsic two-time correlation function
considered in Eq.~(\ref{pfirst}), the correct scaling function for the
aging behavior of the process is given by 
\begin{eqnarray}
f^{\rm intr}(x)\sim f^{\rm meas}(x)~ (1+x)^{\tau-1}.
\end{eqnarray}
Since we are interested in the intrinsic behavior $f^{\rm intr}(x\gg 1)\sim
x^{-r}$, we obtain from our numerical data
\begin{eqnarray}
f^{\rm meas}(x)\sim x^{-(r+\tau-1)}.
\label{imrel}
\end{eqnarray}

Of course, for the random walk near the absorbing wall it is
$\tau=3/2$ from the familiar first-passage time \cite{Hughes}. Thus,
even after correcting for the effect of disappearing walkers, the
intrinsic process still shows aging behavior: The measurable aging effect 
derived in Eq.~(\ref{cabso}), $f(x)=f^{\rm meas}(x)\sim x^{-1}$, leads 
to  an intrinsic aging behavior of $f^{\rm intr}(x\gg 1)\sim x^{-r}$ 
with $r=1/2$ according to Eq.~(\ref{imrel}).
Interestingly, the infinite walk may never touch the absorbing potential
on the wall but clearly feels its effect.

\section{Aging in Self-Organized Critical Models}

The Bak-Sneppen model \cite{B+S} has been studied intensely and with
great numerical accuracy in recent years. We refer to
Ref.~\cite{scaling} for a review of its many features and simply
utilize those facts here.  The model consists of random numbers
$\lambda_i$ between 0 and 1, each occupying a site $i$ on a
$d$-dimensional lattice. At each update step, the smallest random number
$\lambda_{min}(t)$ is located.  That site as well as its $2d$ nearest
neighbors each get new random numbers drawn independently from a flat
distribution between zero and one.  The system evolves to a SOC state
where almost all numbers have values above $\lambda_{\rm c}$, with
$\lambda_{\rm c}$ avalanches formed by the remaining numbers below.

The multi-trade model \cite{BoPa1,BoPa2} is a variant of the
Bak-Sneppen model that provides a series of exact results for the
spatio-temporal correlations in the avalanche process. Especially, an
equation of motion can be derived and solved to obtain a propagator for the
spread of avalanche activity, and to obtain a complete set of scaling
exponents that verify scaling relations previously proposed for
the Bak-Sneppen model. In this model each lattice site is occupied by
$M$ independent random numbers. On each update again the smallest
number in the whole system is located and updated, and one randomly
chosen number 
(out of $M$) from each of the neighboring sites is updated as
well. While the mechanism proceeds in the same way for all $M$ as for
$M=1$ (the Bak-Sneppen model), it can be treated analytically for
$M=\infty$.  

A quantity similar to the two-time autocorrelation function defined in
Eq.~(\ref{twotime}) can be measured for avalanches in the
self-organized critical state of both models.  We focus
on a simple quantity, $P_{\rm first}(t)$, measuring the first returns
of the activity to a given site.  A power law distribution for $P_{\rm
first}(t)$ has been measured numerically for a variety of different
SOC models \cite{scaling} by recording all first returns and has been
derived exactly for the multi-trade model \cite{BoPa1}.   Here we
determine the intrinsic probability $P_{\rm first}(t_w; t)$ to return after $t$
time steps to a site that was visited most recently at time $t_w$ from
the beginning of the avalanche.  Thus, to obtain the first-return
probability, we take the age of the avalanche, $t_w$, into
account. While in the stationary state of SOC models the first return
distribution is generally $P_{\rm first}( t)\sim t^{-\tau_{\rm
first}},~( t\to\infty)$, we find that $P_{\rm first}(t_w; t)$ for
both models considered here scales according to Eq.~(\ref{pfirst}) where
the exponent $\tau_{\rm first}$ can be
related to other critical exponents via scaling relations for SOC
\cite{scaling}.  The origin of the intrinsic aging exponent $r$ appears to 
us to be non-trivial, signaling the breaking of time-translational invariance
in the avalanche dynamics. Unlike in the random walk near
a wall, no symmetry is explicitly broken. The Bak-Sneppen mechanism in
both models evolves on an isotropic lattice with update rules that do
not change with time. The question
then arises whether the exponent $r$ in Eq.~(\ref{pfirst}) can be
related to the known universal coefficients of the stationary SOC
process, or whether it describes new physics in avalanche dynamics of
the Bak-Sneppen model.

\subsection{Numerical Procedure}

In our simulations for both models we have used the equivalent branching process
\cite{PMB}  to eliminate any finite-size
effects.  Initially, at time $t_a=0$, the smallest threshold value is
set equal to $\lambda_{\rm c}$ to start a $\lambda_{\rm c}$
avalanche. In every update $t_a\to t_a+1$, only the signal
$\lambda_{\rm min}(t_a)$ and its $2d$ nearest-neighbor sites receive new
threshold values.  At any time, we store only those threshold values
$\lambda_i<\lambda_{\rm c}$ that are part of the avalanche because
only those numbers can contribute to the signal, i. e. can ever become
smallest number. In addition, we keep a dynamic list of every site
that has ever held the signal at some time to determine the
first-return probabilities. (Since the Bak-Sneppen mechanism for $d<4$ is a fractal
renewal process, activity always returns to a site unless the
avalanche dies.) Avalanches die when there are no
$\lambda_i<\lambda_{\rm c}$ or are stopped at a cut-off $t_a=t_{\rm
co}$, and a new (independent) avalanche is
initiated with $t_a=0$.

At each update $t_a$ we determine the previous time $t_w$ when the
signal was on the same site most recently (if ever). Then its
first-return time is give by $ t=t_a-t_w$, and we bin histograms
labeled by $i=\lceil{1\over 3}\log_2 t_w \rceil$ and $j=\lceil\log_2
t\rceil$.  The data is binned logarithmically so that in each bin a
comparable number of events is averaged over: for each increment
of $j$, the width of the bins for $t$ increases by a factor of $2$,
while for each increment of $i$ the $t_w$ bins increase by a factor of
$8$.  We then normalize for each value of $i$ separately to obtain
the measured first-return probabilities $P_{\rm first}^{\rm meas}(t_w;
t)$. This data is plotted in Figs.~\ref{bsd1norm},\ref{bsd2norm},
\ref{mtd1norm}, and \ref{mtd2norm}.

In each Figure, each graph refers to a different value of $i$,
increasing by a 
factor of $8$ each time from left to right. Each graph possesses two
distinct power law regimes, separated by a crossover.  To determine
the form of the scaling function $f(x)$ for these graphs according to
Eqs.~(\ref{pfirst}), we note that the crossover appears to scale
linearly with $t_w$ in all cases. Thus, we plot 
\begin{eqnarray} 
f^{\rm meas}(x)\sim t^{\tau_{\rm first}} P_{\rm first}^{\rm meas}(t_w;t), \qquad x={ t\over
t_w},
\label{c1} 
\end{eqnarray} 
using the appropriate values of $\tau_{\rm first}$. 
In each case, the data collapses reasonably well onto a
single curve, $f(x)$, which is constant for small argument, and
appears to fall like a power law, see
{}Figs.~\ref{bsd1scal},\ref{bsd2scal},\ref{mtd1scal}, and 
\ref{mtd2scal}.  The exponent of the power law is
given by $r+\tau-1$ according to the relation between measured and
intrinsic data discussed in Sec.~III [see Eq.~(\ref{imrel})]. The
values of $\tau$ are given in Ref.~\cite{scaling} for the Bak-Sneppen
model in $d=1$ and $2$, and $\tau=3/2$ in any dimension for the multi-trade
model.

\subsection{Results for the Bak-Sneppen Model in $d=1$}

We have simulated the Bak-Sneppen branching process in $d=1$ with
$\lambda_{\rm c}=0.66702$, summing over a sequence of all avalanches
up to a cut-off at $t_{\rm co}=2^{27}$. That data consists of a total
of about $10^{11}$ updates. (The results reported here are consistent
with but substantially better than those reported previously in
Ref.~\cite{BoPa3} where  data from avalanches that
did not reach the cut-off was discarded to avoid confusion about the relation
between measured and intrinsic properties discussed in Sec. III.)

\begin{figure}
\epsfxsize=2.2truein
\hskip 0.15truein\epsffile{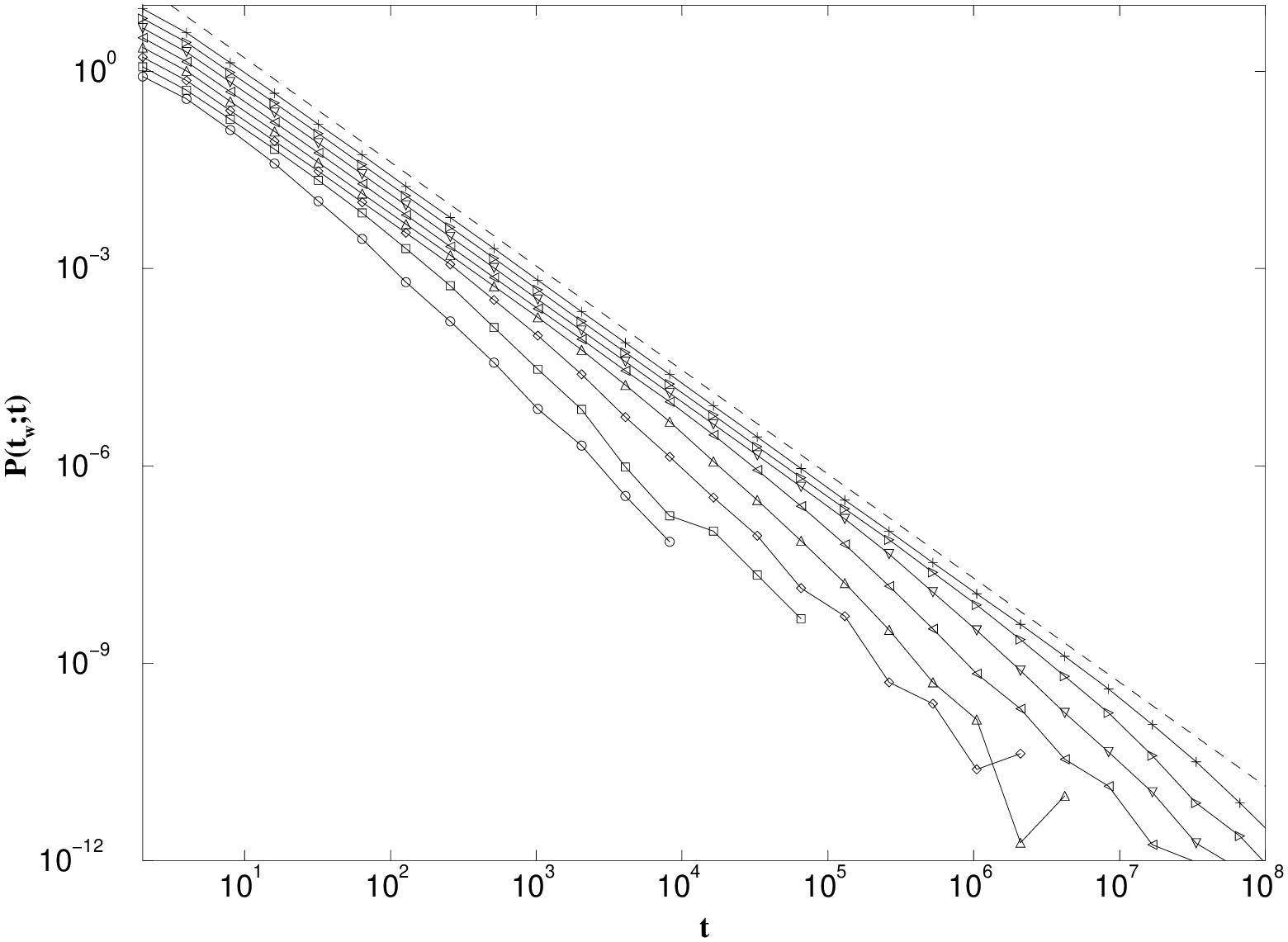}
\caption{\protect\label{bsd1norm}
\narrowtext
Plot of $P_{\rm first}^{\rm meas}(t_w;t)$ as a function of $t$ for the
Bak-Sneppen model in $d=1$. Each graph is 
offset by a factor to avoid overlaps.
}
\end{figure}
\begin{figure}
\epsfxsize=2.2truein
\hskip 0.15truein\epsffile{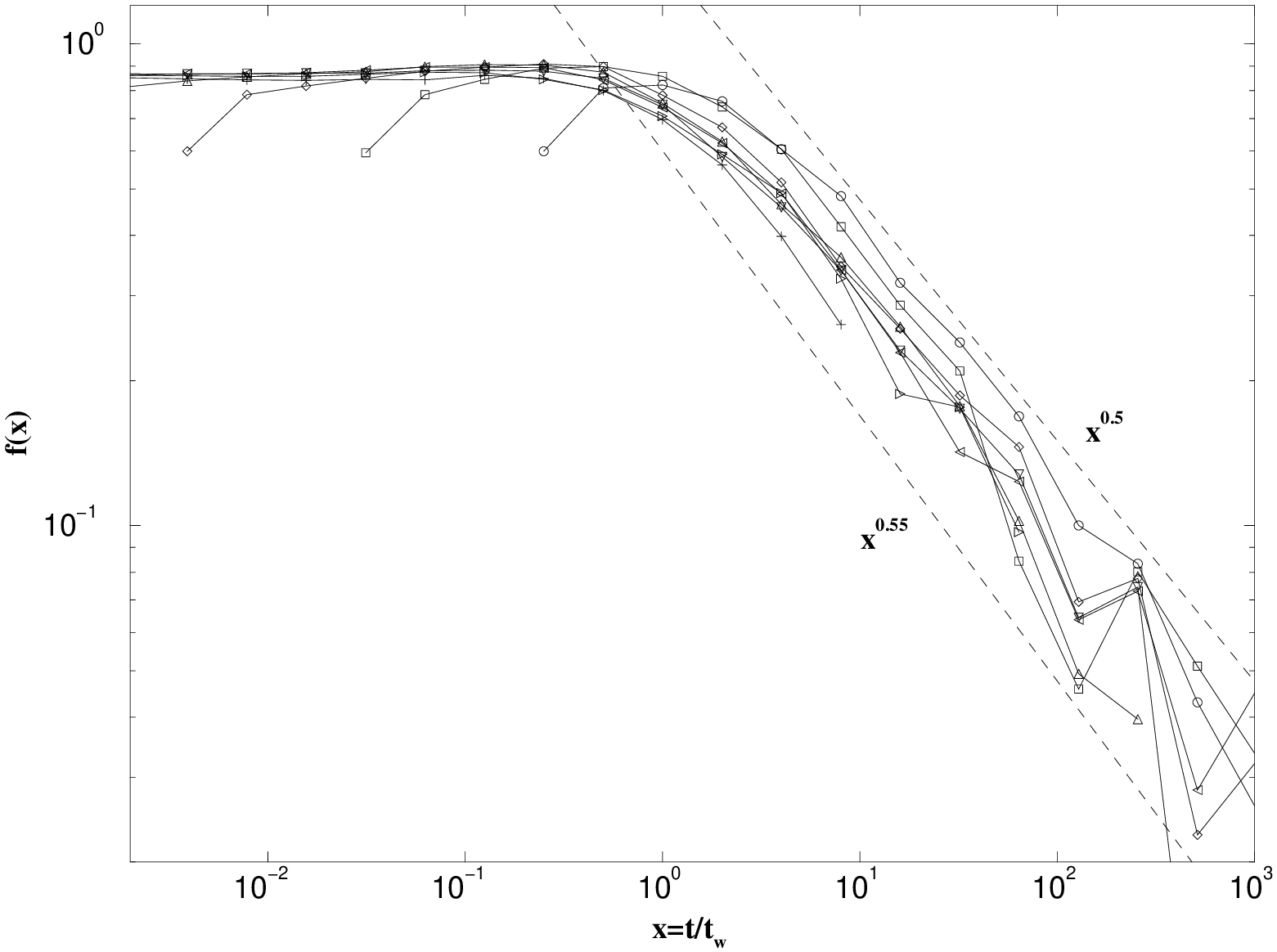}
\caption{\protect\label{bsd1scal}
\narrowtext
Scaling plot for $f^{\rm meas}(x)$ as a function of $x=t/t_w$ according
to Eq.~(\protect\ref{c1}) for the normalized distribution of
the $1d$ Bak-Sneppen model in Fig.~\protect\ref{bsd1norm}.
}
\end{figure}

The distributions for $P_{\rm first}^{\rm meas}(t_w;t)$ as a function of $t$,
normalized for each value of $i$, $8^{i-1}\leq t_w<8^i$, are 
plotted in Fig.~\ref{bsd1norm} for $i=1,\ldots ,8$.
Each graph shows two scaling regimes
separated by a crossover that appears to scale linearly with the
associated value of $t_w$. The initial regime scales with the familiar
exponent $\tau_{\rm first}=1.58$ \cite{scaling} (as indicated by the
dashed line to the right). Cut-off effects become apparent at about
$t\approx 10^7=10\% t_{\rm co}$.

In Fig.~\ref{bsd1scal} we have combined the data into a scaling plot
for $f^{\rm meas}(x)$ according to Eq.~(\ref{c1}) as a function of the scaling
variable $x=t/t_w$. For $x<1$ that data indeed collapses onto a
constant, while we observe a collapse onto a power law over three
orders of magnitude for $x>1$. Deviations from this behavior is
generally due to short-time, transient behavior for $x<1$, and due to
statistical noise deep in the tail of the distribution for $x>1$. We
have bracketed the power-law tail by two dashed lines $\sim
x^{-0.55}$ and $\sim x^{-0.5}$. Thus, we estimate that the exponent of
the tail is given by $r+\tau-1=0.52\pm 0.04$. With $\tau=1.07\pm 0.01$
\cite{scaling}, we finally get $r=0.45\pm 0.05$.

\subsection{Results for the Bak-Sneppen Model in $d=2$}

In this case we have simulated the  with $\lambda_{\rm c}=0.328855$
\cite{scaling}, summing over a sequence of all avalanches 
up to a cut-off at $t_{\rm co}=2^{25}$ (longer avalanche are less
common here than in the $d=1$ case).  We have run about $3\times
10^{10}$ updates for this model.

\begin{figure}
\epsfxsize=2.2truein
\hskip 0.15truein\epsffile{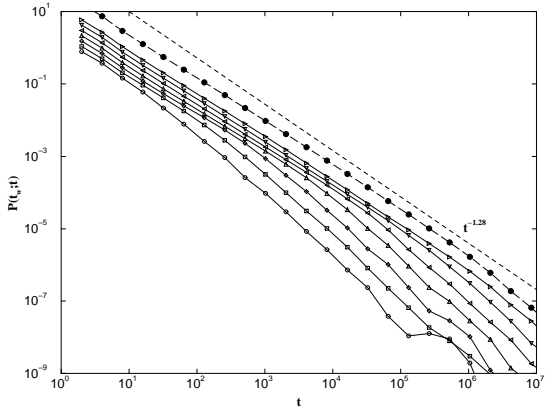}
\caption{\protect\label{bsd2norm}
\narrowtext
Plot of $P_{\rm first}^{\rm meas}(t_w;t)$ as a function of $t$ for the
Bak-Sneppen model in $d=2$. Each graph is 
offset by a factor to avoid overlaps.
}
\end{figure}

The distributions for $P_{\rm first}^{\rm meas}(t_w;t)$ as a function of $t$, again
normalized for each value of $i$, $8^{i-1}\leq t_w<8^i$, are plotted in
Fig.~\ref{bsd2norm} for $i=1,\ldots ,7$. 
As in the $d=1$ case, each graph shows two
scaling regimes 
separated by a crossover that appears to scale linearly with the
associated value of $t_w$. The initial regime supposed to scale with the 
exponent $\tau_{\rm first}=1.28$ determined from a more extensive
simulation in Ref.~\cite{scaling}. That behavior is given by the
dashed line to the right. But each graph approaches that
asymptotic behavior in its initial scaling regime only very slowly,
indicating strong corrections to scaling in this case. Even the
combined data (the dashed line with square marks), not unlike the
corresponding plot in Ref.~\cite{scaling}, approaches asymptotia very
reluctantly. Furthermore, cut-off effects become apparent at
about $t\approx 10^6=10\% t_{\rm co}$.

\begin{figure}
\epsfxsize=2.2truein
\hskip 0.15truein\epsffile{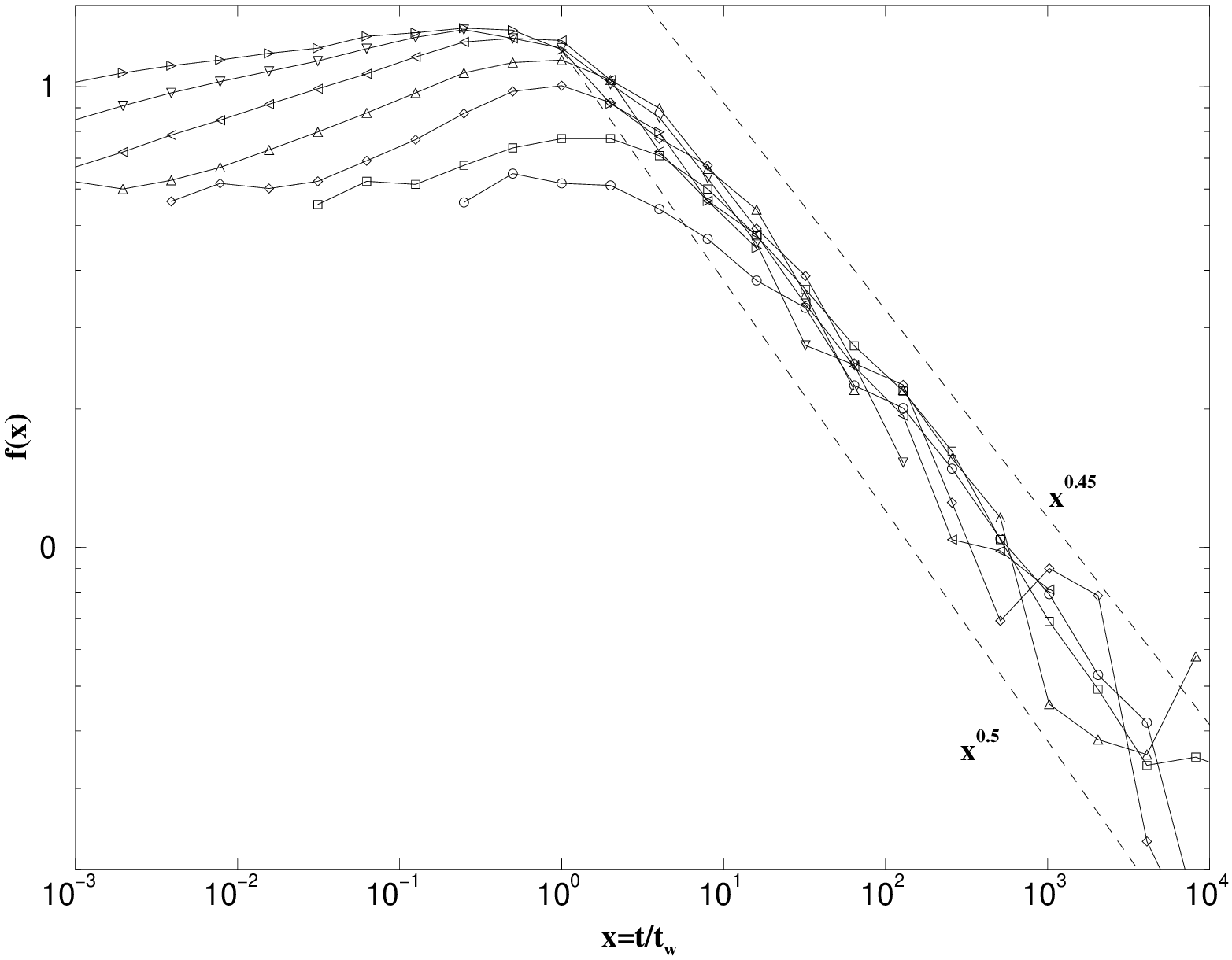}
\caption{\protect\label{bsd2scal}
\narrowtext
Scaling plot for $f^{\rm meas}(x)$ as a function of $x=t/t_w$ according
to Eq.~(\protect\ref{c1}) for the normalized distribution of
the $2d$ Bak-Sneppen model in Fig.~\protect\ref{bsd2norm}.
}
\end{figure}

In Fig.~\ref{bsd2scal} we have combined the data into a scaling plot
for $f^{\rm meas}(x)$ according to Eq.~(\ref{c1}) as a function of the scaling
variable $x=t/t_w$. The collapse onto a constant for the data
at $x<1$ only proceeds slowly due to the aforementioned corrections to
scaling in the first return probability. On the other hand, we observe
a collapse onto a power law over more than three orders of magnitude
for $x>1$ in this case: Since $\tau_{\rm first}=1.28$ is smaller than
in the case $d=1$, many more events occur in the tail of the
distribution. We
have bracketed the power-law tail by two dashed lines $\sim
x^{-0.5}$ and $\sim x^{-0.45}$. Thus, we estimate that the exponent of
the tail is given by $r+\tau-1=0.47\pm 0.04$. With $\tau=1.245\pm 0.010$
\cite{scaling}, we finally get about $r=0.23\pm 0.05$.


\subsection{Results for the Multi-Trade Model in $d=1$}

A number of important properties, like the distributions of the width
and duration of an avalanche and the exponent for the
avalanche dimension ($D=4$), can be derived exactly for this model
\cite{BoPa1,BoPa2}. But many other properties are as of yet elusive or
can only be inferred from scaling relations; for instance, it is 
\begin{eqnarray}
\tau_{\rm first}=2-{d\over 4}.
\label{tf}
\end{eqnarray}
 In particular, no expression for $P_{\rm first}(t_w;t)$
has been obtained so far to study the aging behavior in this model
more explicitly. Thus, as a first step to obtain an explicit expression,
we have simulated this model also in $d=1$ and $2$. The results
suggest that $r=1/4$ in both cases, independent of dimension.

We have simulated the  with $\lambda_{\rm c}=1/2$ since we chose to
update only one number on each neighbor and always replace the minimum
itself with 1 \cite{BoPa1}.  Summing over a sequence of all avalanches 
up to a cut-off at $t_{\rm co}=2^{27}$, we have run about $10^{12}$
updates for this model.

\begin{figure}
\epsfxsize=2.2truein
\hskip 0.15truein\epsffile{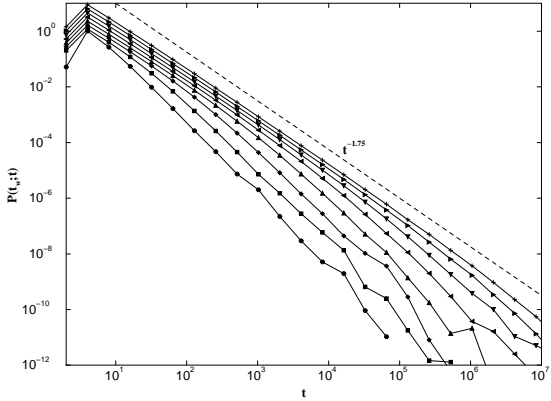}
\caption{\protect\label{mtd1norm}
\narrowtext
Plot of $P_{\rm first}^{\rm meas}(t_w;t)$ as a function of $t$ for the
multi-trade model in $d=1$. Each graph is 
offset by a factor to avoid overlaps.
}
\end{figure}

The distributions for $P_{\rm first}^{\rm meas}(t_w;t)$ as a function of $t$, again
normalized for each value of $i$, $8^{i-1}\leq t_w<8^i$, are plotted
in Fig.~\ref{mtd1norm} for $i=1,\ldots ,8$. 
Each graph shows two scaling regimes 
separated by a crossover that appears to scale linearly with the
associated value of $t_w$. According to Eq.~(\ref{tf}), the initial
regime supposed to scale with the  
exponent $\tau_{\rm first}=7/4$. That behavior is given by the
dashed line to the right. But as in the case of the $2d$ Bak-sneppen
model we observe strong corrections to scaling. Even worse, cut-off
effects already become apparent at about $t\approx 10^6=1\% t_{\rm
co}$. Generally, despite of spending much more time on the simulation,
the data has the poorest quality in this case also because the large
value of $\tau_{\rm first}$ suppresses the occurrence of events in the
long-time tail.

\begin{figure}
\epsfxsize=2.2truein
\hskip 0.15truein\epsffile{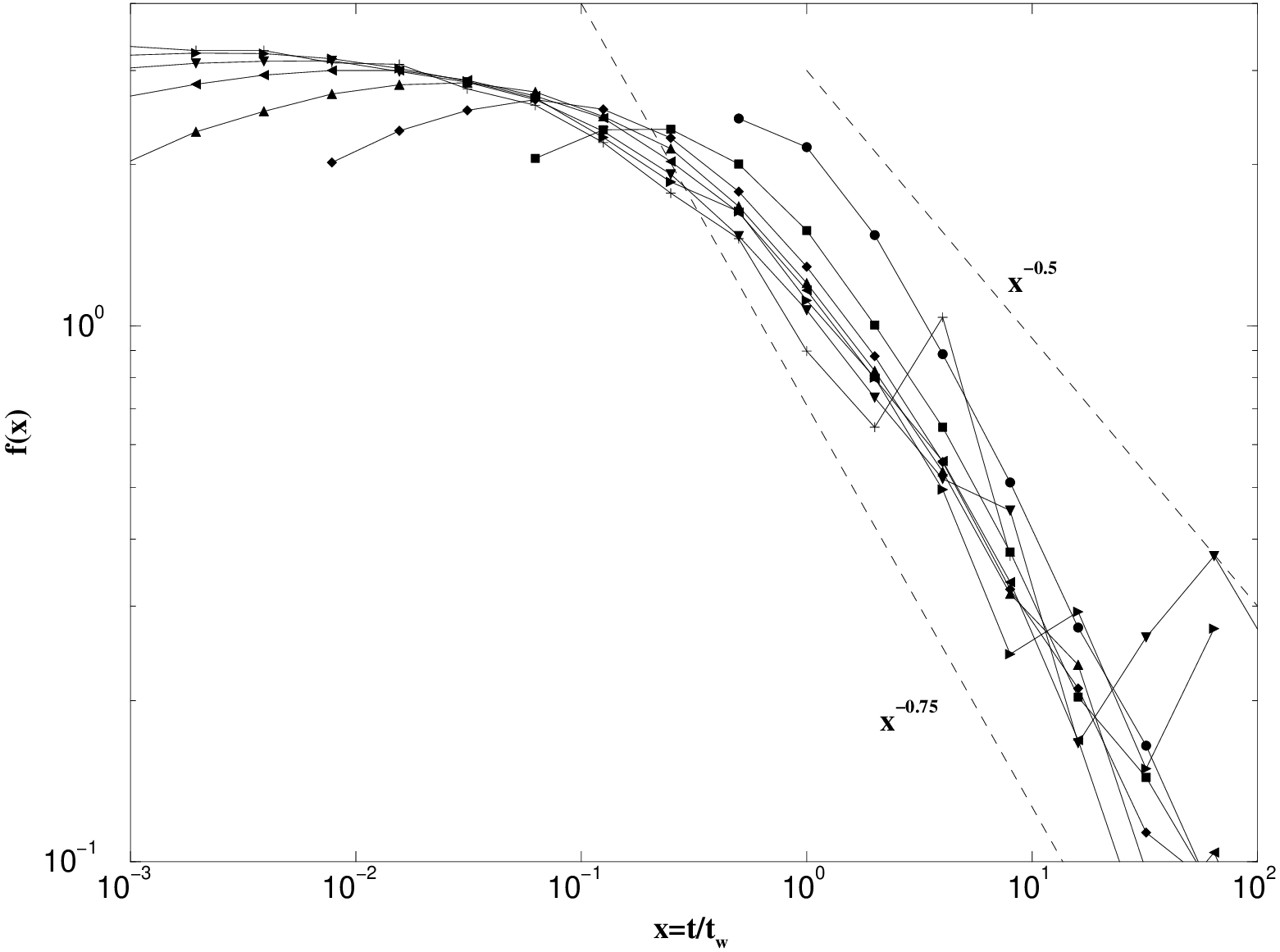}
\caption{\protect\label{mtd1scal}
\narrowtext
Scaling plot for $f^{\rm meas}(x)$ as a function of $x=t/t_w$ according
to Eq.~(\protect\ref{c1}) for the normalized distribution of
the $1d$ multi-trade model in Fig.~\protect\ref{mtd1norm}.
}
\end{figure}

Nontheless, in Fig.~\ref{mtd1scal}, we have combined the data into a
scaling plot 
for $f^{\rm meas}(x)$ according to Eq.~(\ref{c1}) as a function of the scaling
variable $x=t/t_w$. As expected, the corrections to scaling
prevent a satisfactory collapse onto a constant for the data
at $x\ll 1$. Apparently, even the crossover regime is beset by transient
behavior which makes it difficult to localize the transition to the
power-law regime. At best, we can discern scaling over two orders of
magnitude in the tail before the data gets too noisy.  On the other
hand, it is fair to assume from the analytical results that any
exponent in this model should be a multiple of $1/4$. Considering that
the tail clearly scales very close to the dashed line with $x^{-0.75}$ below
and definitely not like the dashed line with $x^{-0.5}$ above, we
believe that $r+\tau-1=3/4$. Thus, with $\tau=3/2$ \cite{BoPa1}, we
conjecture $r=1/4$.


\subsection{Results for the Multi-Trade Model in $d=2$}

We have simulated the  with $\lambda_{\rm c}=1/4$ since we chose to
update only one number on each neighbor and always replace the minimum
itself with 1.  Summing over a sequence of all avalanches 
up to a cut-off at $t_{\rm co}=2^{27}$, we have run about $10^{11}$
updates for this model.   

\begin{figure}
\epsfxsize=2.2truein
\hskip 0.15truein\epsffile{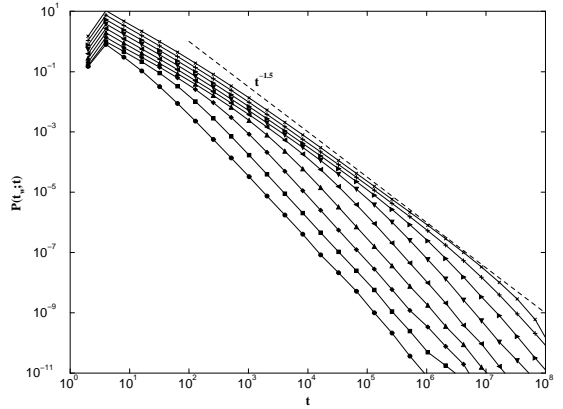}
\caption{\protect\label{mtd2norm}
\narrowtext
Plot of $P_{\rm first}^{\rm meas}(t_w;t)$ as a function of $t$ for the
multi-trade model in $d=2$. Each graph is 
offset by a factor to avoid overlaps.
}
\end{figure}

{
\begin{figure}
\epsfxsize=2.2truein
\hskip 0.15truein\epsffile{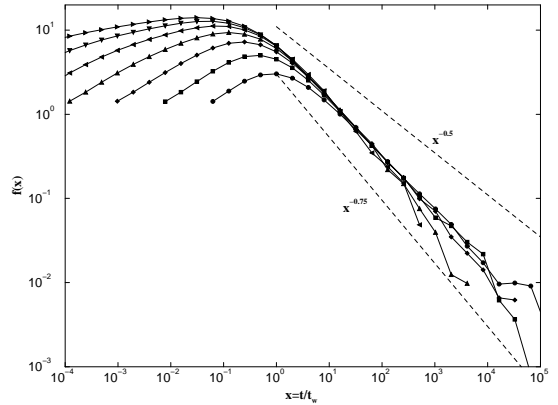}
\caption{\protect\label{mtd2scal}
\narrowtext
Scaling plot for $f^{\rm meas}(x)$ as a function of $x=t/t_w$ according
to Eq.~(\protect\ref{c1}) for the normalized distribution of
the $2d$ multi-trade model in Fig.~\protect\ref{mtd2norm}.
}
\end{figure}

The distributions for $P_{\rm first}^{\rm meas}(t_w;t)$ as a function of $t$, again
normalized for each value of $i$, $8^{i-1}\leq t_w<8^i$, are plotted 
in Fig.~\ref{mtd2norm} for $i=1,\ldots ,9$. 
Each graph shows two scaling regimes 
separated by a crossover that appears to scale linearly with the
associated value of $t_w$. According to Eq.~(\ref{tf}), the initial
regime supposed to scale with the  
exponent $\tau_{\rm first}=3/2$. That behavior is given by the
dashed line to the right. We observe again strong corrections to
scaling and cut-off effects already at about $t\approx 10^6=1\% t_{\rm
co}$.

Combining  the data into a scaling plot again, see Fig.~\ref{mtd2scal},
the corrections to scaling again
prevent a satisfactory collapse onto a constant for the data
at $x\ll 1$. But we observe an excellent collapse of the data in the
tail over roughly three to four orders of magnitude, clearly closer to
the dashed with $x^{-0.75}$ below than the dashed line with $x^{-0.5}$
above. Thus we have $r+\tau-1=3/4$ and $\tau=3/2$ as well in this
case, leading us once more to conjecture $r=1/4$.

\section{Conclusions}

In conclusion, our numerical simulations affirm the existence of a new
and as-of-yet unexplained power law regime in the late time 
behavior of the Bak-Sneppen model that was discover in
Ref.~\cite{BoPa3}. The results underscore the prevalence of
memory effects in the SOC state of this model \cite{BoPa2,Marsili} and
possibly other SOC models \cite{BoPa4}. 

}
We have been able to
rule out a simple relation of the new exponent $r$ to the known
exponents by considering some of the more obvious scaling
arguments (which may not exist at all) \cite{BoPa3}. For instance, we
can consider an avalanche as a random walk near a wall in an abstract random
number space \cite{DeB}. The addition or elimination of a random number
below the threshold $\lambda_{\rm c}$ corresponds to taking a step
away or towards an absorbing wall in a random walk. The avalanche ends
when no random numbers are left below $\lambda_{\rm c}$, i. e. the
walker has been absorbed at the wall. The growth of random numbers
below $\lambda_{\rm c}$ scales like $<n>\sim
t^{d_s}$ for the (intrinsic) infinite avalanche, with $d_s=0.11$ and
$0.25$ for the $1d$ and $2d$ Bak-Sneppen model \cite{scaling} and
$d_s=1/2$ for the multi-trade model in any dimension. If the aging
behavior in these models was due to same effect as in our simple
random walk model in Sec.~II, we would expect that $r=d_s$ which
is inconsistent with the numerical results. It appears that the origin
of the aging behavior is due to more subtle features of the process and
that time-translational invariance might be broken dynamically, as we 
have argued in Ref.~\cite{BoPa3}. The fact that the analytically tractable
multi-trade model also shows nontrivial aging behavior gives us hope
that we will be able eventually to understand its origin.

I am very grateful to Maya Paczuski for discussing the results in this
paper with me.

\end{multicols}
\end{document}